\documentclass[manuscript,screen,authorversion]{acmart}
\setcopyright{none}
\AtBeginDocument{%
  }

\usepackage{amsmath,amsfonts,amsthm}
\usepackage{algorithm,algorithmic}

\usepackage{xspace}
\usepackage{graphicx}
\usepackage{textcomp}
\usepackage{xcolor}

\definecolor{darkgreen}{rgb}{0.0, 0.5, 0.0}
\definecolor{darkred}{rgb}{0.82, 0.1, 0.26}

\newcommand{\para}[1]{\vspace{0.1in} \noindent \textbf{#1}}

\newcounter{challengecounter}
\newcommand{\challenge}{\stepcounter{challengecounter}O\thechallengecounter{}\xspace}

\begin{document}

\title{Software Security Analysis in 2030 and Beyond: A Research Roadmap}

\author{Marcel B{\"o}hme}
\affiliation{%
  \institution{Max Planck Institute for Security and Privacy}
  \country{Germany}
}

\author{Eric Bodden}
\affiliation{%
  \institution{Paderborn University and Fraunhofer IEM}
  \country{Germany}
}

\author{Tevfik Bultan}
\affiliation{%
  \institution{University of California, Santa Barbara}
  \country{USA}
}

\author{Cristian Cadar}
\affiliation{%
  \institution{Imperial College London}
  \country{UK}
}

\author{Yang Liu}
\affiliation{%
  \institution{Nanyang Technological University}
  \country{Singapore}
}

\author{Giuseppe Scanniello}
\affiliation{%
  \institution{University of Salerno}
  \country{Italy}
}

\renewcommand{\shortauthors}{M. B{\"o}hme, E. Bodden, T. Bultan, C. Cadar, Y. Liu, and G. Scanniello}

\begin{abstract}
  As our lives, our businesses, and indeed our world economy become increasingly reliant on the secure operation of many interconnected software systems, the software engineering research community is faced with unprecedented research challenges, but also with exciting new opportunities. In this roadmap paper, we outline our vision of Software Security Analysis for the software systems of the future. Given the recent advances in generative AI, we need new methods to evaluate and maximize the security of code co-written by machines. As our software systems become increasingly heterogeneous, we need practical approaches that work even if some functions are automatically generated, e.g., by deep neural networks. As software systems depend evermore on the software supply chain, we need tools that scale to an entire ecosystem. What kind of vulnerabilities exist in future systems and how do we detect them? When all the shallow bugs are found, how do we discover vulnerabilities hidden deeply in the system? Assuming we cannot find all security flaws, how can we nevertheless protect our system? To answer these questions, we start our research roadmap with a survey of recent advances in software security, then discuss open challenges and opportunities, and conclude with a long-term perspective for the field.

\end{abstract}


\maketitle

\section{Introduction}
Over many years, the primary focus of the software engineering community has been on building larger and more reliable software systems with minimal effort. Today, we are proud to see the world's digital economy run reliably on a super-scale, ever-changing, hyper-connected network of software systems. With the power of the software supply chain, cloud computing resources at our fingertips, and a large automation ecosystem at our disposal, we can build complex systems from existing components at an unprecedented pace. Recent advances in artificial intelligence (AI) promise further automation even of creative tasks (incl. auto-programming to build even larger systems). Our software systems are becoming more heterogeneous and extend further into the real world, as for instance, virtual reality, the Internet of things, autonomous cars, and robots.

Over the coming years, we anticipate that software security analysis will become another important focus of the software engineering SE community. How can we make our large, interconnected software systems robust against attacks? In this paper, we draw a research roadmap towards 2030 and beyond. We identify concrete challenges and opportunities for the security analysis of our software systems of the future and provide specific directions of research.

\section{State Of The Research and Practice}

Software security analysis has seen significant progress over the last decade.
Several program analysis techniques, ranging from static analysis~\cite{csa,infer,coverity} to fuzzing~\cite{afl,libfuzzer,jqf} to symbolic execution~\cite{klee,cute,spf} have reached maturity, and many software systems have started to adopt them.
Combined with other software development changes, such as new-generation IDEs~\cite{vscode,clion}, better code hosting platforms~\cite{github}, and higher adoption of code review and continuous integration~\cite{code-reviews:survey,ci-effects}, this has enabled the development of larger and more complex software systems.

\subsection{Software Analysis Techniques}

This section briefly describes the most widely used software analysis techniques employed in practice, particularly in a security context:

\para{Formal verification} provides the highest level of security guarantees, with formal machine-checked proofs being able to establish the absence of certain classes of bugs.
In the last 15 years, the field has seen tremendous progress, with researchers building several safety and security-critical software systems, such as the seL4 microkernel~\cite{sel4} and the CompCert compiler~\cite{compcert}.
Yet, there are, unfortunately, two main inter-related limitations of formally-verified systems: (1)~they require years of PhD-level expertise, with the specification often larger than the verified code itself, and (2)~the resulting systems lack many of the features and/or performance of their non-verified counterparts. In addition, evolving these systems is often costly, as the verification effort is not always modular.

\para{Static analysis} reasons about program code without executing it. Abstract interpretation~\cite{cousot1996abstract} is a well-known type of static analysis, with sound reasoning, which can also be used for formal verification. Especially in the security domain, more lightweight vulnerability detection techniques have become very popular~\cite{li2017static}. These frequently focus on identifying the most common kinds of vulnerabilities, e.g., the SANS 25~\cite{sans25}. These techniques are usually based on data-flow analysis that---for scalability reasons---oftentimes forgoes complex reasoning about arithmetic and other aspects. To provide a ``good signal'' with few false positives, most such approaches these days also accept a certain level of unsoundness~\cite{pauck2018android}. While recent work has shown that clever program abstractions and algorithms can improve analysis precision and speed at the same time~\cite{spath2019context}, scalability remains a challenge when it comes to analyzing large code bases. Another major shortcoming of static analysis, shared with other techniques, is that analyzers need to be configured: they only report what they are configured to report, which is why they must be told, e.g., which particular API calls in which combination can lead to which kinds of vulnerabilities~\cite{piskachev2019codebase,piskachev2021secucheck}. In other words: static analysis also does not completely forego a specification, yet here one specifies vulnerability types, not program functionality.

\para{Fuzzing}~\cite{fuzzing:ieeesw}, at a high level, involves testing software with randomly generated inputs.
There are three main forms of fuzzing: black-box, grey-box, and white-box fuzzing, with the latter also known as dynamic symbolic execution and discussed separately below.
Black-box fuzzing does not use any knowledge of the implementation, nor any execution feedback other than what can be observed by running the software in a black-box manner. At the same time, it requires domain knowledge to be effective, such as grammar for the inputs accepted by the program.  The technique has been applied with a lot of success in several areas, for instance for optimizing compilers and database management systems (DBMS), where it has found hundreds of critical bugs in mature compiler and DBMS implementations~\cite{csmith,emi,rigger:pqs}.
Greybox fuzzing is the most widely used form of fuzzing for general software systems, which is guided by the code coverage achieved by the generated inputs. It is typically combined with mutation-based input generation. Greybox fuzzing was pioneered by AFL~\cite{afl} and now there are dozens of greybox fuzzers available for many different languages and domains. OSS-Fuzz~\cite{oss-fuzz} is a service provided by Google which runs several state-of-the-art greybox fuzzers on open-source software. To date, it has found over 10K vulnerabilities and over 36k bugs in over one thousand projects.

\para{Dynamic symbolic execution}~\cite{symex:cacm13,exe,cute,spf,sage}, also referred to as \textit{concolic execution}, \textit{whitebox fuzzing} or just \textit{symbolic execution}, is a program analysis technique that can systematically explore paths in a program by using an SMT constraint solver~\cite{smt:cacm11} to reason about path feasibility and generate concrete test inputs for each explored path. While the technique has a long history going back to the 1970s~\cite{king76,clarke76}, it is only in the last couple of decades that it has become practical, enabled by both algorithmic advances and practical tool design~\cite{symex:survey}. Nevertheless, important scalability challenges remain, both in terms of path explosion and constraint solving~\cite{symex:cacm13}. Dynamic symbolic execution has started to be adopted in practice~\cite{symex-practice,klee-2019} by companies such as Fujitsu~\cite{klover}, Microsoft~\cite{sage} and Samsung~\cite{aot}, and nowadays several mature open-source tools exist for different languages and platforms~\cite{klee,spf,symcc,binsec}.

\para{Machine learning} was applied in software security analysis mostly to the problem of further automating code-level vulnerability detection, ranging from approaches that help configure traditional static data-flow analyses~\cite{piskachev2019codebase} to such that try to detect vulnerabilities solely using machine learning, without data-flow analysis~\cite{chakraborty2021deep}. For the latter case, while in cross-validation lab experiments, many approaches in this field at first showed extremely promising results~\cite{devign}, it has been unfortunately proven again and again that with current models the recognition quality quickly degrades when the machine learning models are tested on code on which they had not been trained~\cite{krishnan2023static, gao2019negative}. This result is similar to what has been observed for malware detection~\cite{allix2016empirical}. Another issue with many such classifiers is that they merely report that they suspect \emph{some} vulnerability in a given method or even file---information that is hardly useful to developers and can at best help prioritize further vulnerability detection activities. While more recent approaches attempt to report findings on the line level~\cite{fu2022linevul}, and try to report at least the given CWE classification of the suspected vulnerability, it remains to be shown whether developers can act on such findings.

\para{Runtime protection mechanisms} are used during software deployment, targeting vulnerabilities that development- and testing-time activities have failed to detect. These techniques target specific classes of errors, typically memory vulnerabilities~\cite{sok-memory}, and usually make it harder rather than impossible, for attackers to exploit those vulnerabilities. Well-know techniques include stack canaries~\cite{stackguard}, control-flow integrity~\cite{cfi}, data-flow integrity~\cite{cfi}, code-pointer integrity~\cite{cpi} and write integrity testing~\cite{wit}.

\subsection{Detected Errors}

Given that most of our security and safety-critical code continues to be developed in unsafe languages like C and C++, most attention has been given to memory safety errors~\cite{sok-memory}. On the one hand, techniques targeting such errors have been designed and adopted in practice, such as compiler sanitizers~\cite{asan,msan}, which can find bugs such as buffer overflows and use-after-free errors. On the other hand, techniques that can in principle find a broader class of bugs, such as static analysis and symbolic execution, have been primarily applied to find memory safety errors and injection vulnerabilities, both due to their importance and their easily available oracles~\cite{klee,cute,infer,csa}.

To go beyond memory safety, researchers have exploited differential and metamorphic oracles. For instance, differential testing has been extremely successful in the context of compiler testing~\cite{csmith}, while various types of metamorphic transformations have been effective for database management systems~\cite{rigger:pqs}.

Certain types of functional bugs may escape differential and metamorphic testing. Specifications have been for long proposed as a way to write correct software, but adoption has been limited, particularly in mainstream languages.

\section{Challenges and Opportunities}
As we look forward, we predict a substantial shift in the \emph{key challenges} for security analysis. Not only will our software systems grow larger, more interconnected, more distributed, and evolve faster. We also expect to see a greater dependence on the software supply chain, a deeper integration with the physical world, and a broader range of important security properties beyond memory safety.
There are also many \emph{untapped opportunities} for us to develop the area of software security analysis. For instance, recent advances in machine learning hold a strong promise for non-traditional, empirical approaches for reasoning about the security of a software system \cite{statistical}. Outside of the technical aspects, as the field matures, we should work on \emph{pedagogical aspects}, such as establishing and teaching the foundations of software security, as well as \emph{legal aspects}, such as questions of liability or accountability in the digital world.

For easy reference, each challenge or opportunity is identified uniquely using a counter and a name. For instance, \emph{\challenge.~This is a challenge or opportunity}.\addtocounter{challengecounter}{-1}

\subsection{Overview: Software Systems of the Future}
Over the last several decades, software engineering has mainly focused on the \emph{design and implementation} of the software systems that underpin our digital world, including our digital economy, our digital entertainment and social networks, as well as our apps and devices that we use every day. Going forward, we expect, the focus will slowly move to the \emph{evolution, maintenance, and hardening} of these software systems. Once the software supply chain is set up during design and implementation, even if the software system itself is properly maintained, it is important also to monitor and maintain the integrity of its supply chain. This will require \emph{ecosystem-scale analyses} of large networks of dependencies.

Software systems will compose more dynamically and work in concert with many other systems that are in many cases unknown at development-time. Such systems of systems will be \emph{highly interconnected and distributed} to an arbitrary complexity. Individual nodes may be elastically added or removed at any time depending on the current user-demand or compute-availability. The dynamic organization of computation makes it difficult to analyze the security of such systems a-priori. The distributed nature of computation poses challenges for the confidentiality and integrity of the data that is being processed.

Software systems will change more rapidly. Many software systems are built and maintained with a continuous integration / continuous deployment (CI/CD) pipeline enabled which allows for a rapid testing and deployment of new features, bug fixes, or security patches. With the advent of effective tools for (semi-)automated programming \cite{autocoderover} and vulnerability remediation, we expect the rate of change to increase further. Submitted changes, bug fixes, and security patches may be produced or reviewed entirely mechanically \cite{52878}. This increasing level of automation in the development process poses interesting opportunities for future work in automatic software security analysis.

Software systems will be more heterogeneous and potentially integrating machine learning components. Some components may be written in a memory-safe language while others are not. Some components may be implemented in a type-safe language while another may not. Some components may be realized as machine learning model that returns approximate, probabilistic results while the remainder of the system may be entirely deterministic (modulo a thread schedule). An effective software security analysis tool should be able to deal with such heterogeneity.

Software systems will be embedded further into our physical world. Virtual reality systems will project virtual objects into our physical space for us to interact with. Embodied AI will turn virtual chat bots into embodied robot assistants that interact with the real world to test their hypotheses about the real world. More devices will be connected to the internet (via fog and cloud computing). Cars receive software updates over the air. Transport is increasingly automated. While rich with opportunities, from a safety and security perspective, the increasing ability to interact with the physical world also poses increased risks of physical damage up to loss of life.

\subsection{Machine Learning and Security}

Recent advances in machine learning promise a paradigm shift in software security analysis.

\subsubsection{Security for Machine Learning (Sec4ML)}
Traditional software systems are \emph{programmed}, i.e., composed of human-written instructions for the machine to process inputs. So, most existing security analyses work by processing these instructions to reason about the security of the software system.
However, we expect that future software systems, at least in part, will be \emph{learned}. Machine learning methods, such as deep learning and large language models have proven extremely useful in the automation of certain tasks. For instance, rather than humans writing programs that provide precise instructions on how to generate a video, an image, or even source code from a text description (called prompt), it might be sufficient to train a machine learning model with large amounts of data.

\para{\challenge. Analysis of ML systems}. How can we make statements about the properties of a program whose behavior emerges from a network of neurons with real-valued weights? From a \emph{formal perspective}, we could investigate transpilation techniques \cite{dnnverif} which turn a machine-learning model into a traditional program (i)~that is guaranteed to exhibit all and only those behaviors as the original model, and (ii)~that can be analyzed by existing software verification or static analysis tools. Skipping the intermediary traditional program, we could develop new analysis techniques that directly extract a formal model of computation representing all executions of that model and allow us to formally reason about the properties of that model. From an \emph{empirical perspective}, we could investigate statistical methods or PAC learnability
 to quantify our uncertainty about empirical statements about a model's properties, given a sample from the operation distribution of inputs for that model \cite{reachability}. We could employ experimental methods to make empirical statements about hyperproperties \cite{hyperproperties} of an ML-based software system (e.g., robustness quantifies the degree to which a small change in the input causes a change in the output of the model).

\para{\challenge. Vulnerability types in ML systems}. Which security flaws might exist in ML-based software systems that could be exploited with malicious intent, and how do we analyze and mitigate them? We already know about certain broad types of attacks on machine learning systems. In a data poisoning attack, the malicious actor manipulates the training data to influence the behavior of the trained software system that is eventually deployed. In a data privacy attack (e.g., membership inference, model extraction, or model inversion), a malicious actor can extract sensitive data from the deployed system. In a model evasion attack, a malicious actor learns to manipulate the deployed system's outcome, e.g., to evade detection. There are most certainly other types of vulnerabilities; some may be domain- or model-specific. How can we detect and mitigate them?

\para{\challenge. Vulnerabilities in ML-generated code}. Which security flaws are typically observed in ML-generated code and how do we detect them?
How do we improve the security of ML-generated code?
Can ML-generated code be more secure than human-written code?
For little over a year, we have seen increasing adoption of ML-based programming assistants that can translate simple comments \cite{copilot,devign} or complex Github Issues \cite{autocoderover} into source code written in any programming language. Such ML assistants free developers from resolving mundane issues and help them focus on the high-level, creative part of the development process. However, these machine programmers are also known to ``hallucinate'' \cite{insecureLLM,insecureLLM2}.
Today, they tend to generate faulty and potentially insecure implementations. If the complicated, auto-generated code is accepted by the busy programmer without much reflection, ML-generated vulnerabilities might start to sneak into production code at an increasing velocity.

\subsubsection{Machine Learning for Security (ML4Sec)}
Machine learning has always played an important role in software security analysis, e.g., for intrusion detection \cite{intrusion}, malware analysis \cite{malware}, and vulnerability detection \cite{vuldeepecker}. Specifically, machine learning is used to identify patterns that are \emph{correlated} with security flaws. While results on academic benchmarks seem promising, other factors, like overfitting or spurious correlations, cannot be excluded as alternative explanations for the required code reasoning capabilities \cite{arp2022dos,chakraborty2022survey,risse2023limits}. However, recent developments in large language models, particularly towards systematic reasoning and planning \cite{chainofthought}, and further developments in causal reasoning from data \cite{causalinference} hold exciting prospects for the application of machine learning to software security analysis.

\para{\challenge. ML4SbD} (Secure-by-Design). We anticipate that multi-model machine-learning models, which, for instance, can work with not just text but also drawings, will reshape very soon the way in which we design and document software systems. One can easily envision systems that derive software architectures in a joint interaction with developers. But how can we make sure that ML models, if used in this way, come up with ``the right'' design? How can we factor in security? Will it be problematic if such models yield non-deterministic output? And can machine learning even help detect security flaws on the architectural level?

\para{\challenge. ML4VD} (Vulnerability Detection). How can we use recent advances in ML to swiftly analyze source code written in arbitrary languages, at an arbitrary scale, with a tolerable number of false positives or negatives?
Existing approaches that work without executing the program require some human-provided encoding of the pertinent semantic rules of the programming language. Machine learning can do away entirely with such semantic rules and---given enough training data---works for programs written in any language. For instance, \emph{defect prediction} \cite{defect} has been developed over many years as an ML-based approach to identify potentially defective components via code property correlates. Deep-learning based methods require substantial training data, and it is an open question if such data can at all be properly curated in sufficient amounts for these models to work effectively.  However, recent advances in deep learning (DL), large language models (LLMs), and statistical methods \cite{statistical,reachability} promise an entirely new form of reasoning over the properties of a program. As LLMs are known to make up facts (``hallucinate''), we should explore the limits of these approaches and identify where they work well and where they do not. Given the enduring success of existing formal/symbolic approaches in program analysis, we further envision the development of new neurosymbolic approaches.

\para{\challenge. Tool configuration or assistance}. How can we reduce the level of expert knowledge required to set up security tooling for a given software system most effectively? For interactive security tooling, how can we increase automation and further assist or even replace the engineer with ML? Most security tools are designed to be general, i.e., to work for any software system that is in scope. To adapt such a tool for a specific system, it needs to be \emph{configured}. For instance, to attach a fuzzer to the system the user needs to write the required fuzz drivers \cite{ossfuzz} and provide the required input format or protocol \cite{aflsmart,snapfuzz}. To enable a static analysis security tool (SAST), the user needs to add it to the Continuous Integration (CI) pipeline and possibly to the build process. Such a setup requires expert knowledge about both, the security tools as well as the software system. The recent success of LLMs in similar high-level, creative tasks (e.g., translating hundreds of pages of natural language protocol specification into a format that is usable by a fuzzer \cite{chatafl}) is a promising precedent. In the future, we envision that even the layman can set up security tooling for any specific software system.

\para{\challenge. Sound evaluation}. How can we properly evaluate the true capabilities and limits of ML-based security tooling to discover or mitigate security flaws in important software systems? How can we establish a level playing ground for a sound empirical comparison to other security analysis approaches, like static analysis? In ML-based vulnerability detection, some papers report a performance on public benchmarks that greatly exceeds all expectations, given the general experience in the industry with other security tools (e.g., lt. 6\% false positives; 7\% false negatives \cite{vuldeepecker}). Yet, they cannot distinguish between vulnerable and patched functions \cite{risse2023limits}. Going forward, we need to study the reasons for those outstanding results, identify concrete benchmarking pitfalls, and develop reliable methodologies to evaluate ML techniques for security analysis that systematically exclude alternative explanations for the observed performance.

\subsection{Security of Evolving Software}

Software systems have always been in constant change, but their evolution is due to further acceleration as more tools are being integrated into the development process and AI-based systems are becoming capable of repairing code and contributing new features. As a result, program analysis tools will need to become more agile in the way they are deployed, focusing on analyzing the recently introduced code changes rather than on the overall system.

\para{\challenge. Program analysis for fast-evolving software.} How do we design program analysis techniques that keep up with the high evolution of modern software systems? Preliminary techniques have already been proposed in the literature~\cite{katch,aflgo}, but they still lag behind their whole-program counterparts. An important challenge is that such techniques have to be fast, so as to not slow down software development; for instance, while whole-program fuzzing campaigns are often expected to take 24h~\cite{fuzz-eval}, incremental ones should take on the order of minutes~\cite{fuzz-ci}. How can incremental updates~\cite{arzt2014reviser} be used to speed up static analysis?
Program analysis techniques should take advantage of the runs performed on earlier versions~\cite{memoized-symex,moklee}, as well as use the behavior of the previous version as an implicit oracle~\cite{shadow}.

\subsection{Supply Chain Security}

Yesterday's software systems were developed and analyzed as monoliths that were often entirely developed by a single vendor.
Today’s software systems are (recursively) composed of many third-party components. \emph{This is the software supply chain}. The security posture of the projects behind these third-party components may drastically vary. Some projects may have a vulnerability disclosure policy and track unique identifiers (CVEs) for the software vulnerabilities that were present across different versions of the project. Others may not be actively maintained anymore and are riddled with known vulnerabilities that are so automatically included in the complete software system.

Tomorrow's software systems will contain new vulnerabilities that could be introduced into any part of the supply chain at any time, e.g., by reusing code snippets from Q\&A forums or source-code forges, that include both known and unknown vulnerabilities. Some projects, like the Linux kernel,\footnote{``[..] the CVE assignment team is overly cautious and assigns CVE numbers to any bugfix that they identify''; \url{https://docs.kernel.org/process/cve.html}.} may have a liberal CVE-assignment policy while others never request CVEs for their security flaws which renders the number of CVEs in the supply chain of a software system an unreliable measure of its security posture. Moreover, with the increasing recent adoption of LLMs for the generation of code, security issues hidden in third-party artifacts could be reused and imported in much more flexible ways, which also challenges the existing infrastructure of supply chain security.

\subsubsection{Software Composition Analysis}
A software composition analysis (SCA)~\cite{10.1145/3475716.3475769, 9206429} automatically identifies the third-party components of a software system to detect known vulnerabilities, license compliance issues, or other risks or quality issues arising from the usage of those third-party components. Existing work on SCA has focused on code clone detection either in the system's source code~\cite{golubev2021multi,hung2020cppcd} or the distributed binary~\cite{li2017libd,zhan2021atvhunter}.

\para{\challenge. SCA in the era of ML-generated code}. How can we reliably identify code snippets that are copied from a code repository that is license-protected or that is potentially riddled with security flaws? If the adoption of AI-assisted programming tools continues at the current rate, an increasing proportion of code added to our code bases will be ML-generated. As these tools pull from a vast array of sources to generate or suggest code snippets, accurately tracking and analyzing the security of these piecemeal components becomes daunting~\cite{pearce2022asleep, asare2023github, sandoval2023lost}. This fragmentation complicates dependency and vulnerability mapping in SCA, making it difficult to ensure comprehensive coverage in vulnerability scans. Furthermore, the fragmented code reuse also introduces significant challenges for the identification of plagiarism of existing artifacts~\cite{sun2022coprotector,yu2023codeipprompt,li2023protecting}, which further complicates the copyright and license detection for SCA. To tackle these challenges, we require new SCA techniques at the level of code snippets, like software genes~\cite{wu2023software}, for improved supply chain security assurance.

\para{\challenge. Risk analysis}.
 How can we evaluate the impact of security flaws in third-party components on the host software system with reasonable accuracy and scalability?
Even if SCA tools reliably identify the third-party components with known security flaws (i.e., n-day vulnerabilities), it is left to the user to confirm whether such security flaws or their interactions yield a vulnerability of the host system. A variety of static analysis techniques have been developed to automate this process and identify specific vulnerable execution paths~\cite{jam2021,zhang2023mitigating,wu2023understanding} or to determine the host's general security posture w.r.t. the exploitation of n-day vulnerabilities~\cite{fang2020fastembed,10.1145/3460120.3484594}. However, there is a notable lack of dynamic verification techniques to generate a witness host execution to confirm a vulnerability, and existing static analysis techniques are often either not scalable or too imprecise.

\para{\challenge. Tool support across the software development life cycle} (SDLC). How can we support the automation of supply chain security analysis across the different phases of the SDLC of a project and its dependencies? Similar to the supply chain in manufacturing, the software supply chain consists of people, processes, tools, third-party components, and other artifacts that play a role in the development and maintenance of a system~\cite{supplychain}. Throughout a project SDLC, we can identify several threats to supply chain security. During \emph{development}, developers could submit code containing security flaws~\cite{9402087, guo2023empirical} or add vulnerable third-party
dependencies~\cite{pashchenko2020qualitative} either accidentally or via compromised accounts or malicious insiders \cite{gong2019detecting}. During \emph{build and deployment}, tools such as the compiler or build server may be compromised and inject security flaws~\cite{trustingtrust,compiler-backdoor:blog15}. During \emph{distribution}, package managers and other registries could distribute malicious third-party components, e.g., via typosquatting~\cite{taylor2020defending}, compromised maintainer accounts~\cite{zahan2022weak}, exploits of dependency resolution mechanisms~\cite{gu2023investigating}, or the persistence of outdated dependencies~\cite{zhang2023mitigating, hu2023empirical}. During \emph{maintenance}, the security posture of a project may degrade over time when maintainers move on to other projects or simply do not have the time anymore. This leaves downstream users exposed to known security flaws until they are eventually fixed, if ever~\cite{stringer2020technical}. All of these challenges require new techniques to support the security of a software system across the entire SDLC.

\subsubsection{Supply Chain Ecosystem}
Third-party components of a software system are often distributed via package managers (PMs) and reused within an ecosystem, constituting a network of dependencies. Examples of PMs and the corresponding ecosystems are NPM/JavaScript~\cite{liu2022demystifying}, Maven/Java~\cite{10.1145/3475716.3475769}, PIP/Python~\cite{vu2020typosquatting}, GoLang~\cite{hu2023empirical}, and Android~\cite{zhan2021atvhunter}.
Apart from third-party libraries, pre-training models, datasets, and cloud services are also critical components in new emerging systems, such as LLM-based systems, distributed systems, cloud platforms, and cyber-physical systems. These new types of dependencies also become potential entry point for attacks.

\para{\challenge. Longitudinal studies of ecosystem security health}. How can we detect long-term or emerging security threats in a swiftly evolving network of software dependencies involving multiple component versions? How does the overall security health of an ecosystem evolve over time? Existing studies of such ecosystems have focused on the identification of security threats \cite{zimmermann2019small} and the propagation of vulnerabilities \cite{liu2022demystifying}. As these ecosystems evolve and new ecosystems emerge, in the future similar studies should be repeated in regular intervals and summarized longitudinally. We should develop the capabilities to identify and track various types of attacks on an ecosystem and to monitor how the security posture of individual projects impacts the overall security health of the entire dependency network.

\para{\challenge. Ecosystem-wide vulnerability remediation}. How can we improve the security of a software system that depends on third-party components with known vulnerabilities? How can we improve the overall security health of an ecosystem?
Some projects in an ecosystem might reach their end-of-life, for one reason or another~\cite{huang2022characterizing}, but even if every project was very well maintained where security flaws are fixed as soon as they are found, there is still technical lag in the propagation of these fixes to its dependants~\cite{gonzalez2020characterizing}. Some hosts (i.e., dependants) might be reluctant to update their otherwise trustworthy dependencies and still use an older version~\cite{mirage}. Hence, vulnerable versions should be identified~\cite{bao2022v}, vulnerability patches should be back-ported to these versions~\cite{decan2021back}, and affected dependants should be identified and updated~\cite{chen2020automated}.
To minimize the risks of breaking updates, multi-version execution techniques can be used to roll back any failing updates~\cite{mx:icse13,mvedsua:asplos19}.
To minimize the reliance on (vulnerability-inducing) third-party components, developers can use debloating techniques to trim redundant dependencies from a software system~\cite{bruce2020jshrink,wangyingdebloating}. If the system depends on a third-party component with known vulnerabilities, existing SCA tools would suggest adopting the corresponding patch~\cite{roumani2021patching}, to update the third-party component (where the vulnerability is fixed)~\cite{wang2023plumber}, or to migrate to a different component that implements the required functionality~\cite{he2021multi}.

However, most existing remediation strategies are vulnerability-specific with little adjustment to the host where the remediation is applied. In other words, remediation might lead to potentially breaking changes in the host itself~\cite{zhang2023compatible} or the host's dependants~\cite{10.1145/3551349.3556956}.
Moreover, beyond remediation for individual dependants, we should develop ecosystem-wide intervention strategies to maximize the overall health of the ecosystem in the presence of known vulnerabilities or emerging security threats.

\para{\challenge. The software supply chain of emerging systems.}
How can we identify and properly manage new types of dependencies in emerging systems?
How can we systematically identify the new attack surfaces alongside?
How can we propose comprehensive detection and management solutions to mitigate these newly emerging threats?
The software supply chain of emerging systems faces significant challenges due to its complex, multi-layered nature. Managing dependencies, particularly with pre-trained models, and cloud services, beyond third-party libraries, is increasingly difficult, often resulting in opaque, hard-to-track components. Security vulnerabilities from external services~\cite{verma2020cloud,parast2022cloud}, becomes a potential entry point for attacks, especially in distributed and cloud architectures. The rapid and untraceable updates~\cite{yasmin2020first,golmohammadi2023testing} of external services could introduce potential unreliable functionalities, which should be further included into the management of supply chain.
Ensuring transparency and traceability across these diverse systems, particularly in ML and LLM-based models, is another critical challenge. Pre-training models~\cite{jiang2022empirical,guo2022threats,yao2024survey, zhang2023red,feng2023detecting}, training frameworks~\cite{zhao2024models,wu2024new,halawi2024covert}, as well as poisoned datasets~\cite{cotroneo2024vulnerabilities, pathmanathan2024poisoning,bowen2024scaling}, are also vital components in supply chain, proper detection and management of corresponding threats should also be concerned.
Additionally, cyber-physical systems, such as IoT systems, introduce heightened risks due to their integration of software with safety-critical physical components~\cite{maurya2020reliability}, their supply chains are intricate systems that involve the production and integration of not only software but also hardware components~\cite{ben2019internet, manavalan2019review}, their compliance with safety-critical standards for industries, such as healthcare~\cite{lopez2023comprehensive,cartwright2023elephant, khinvasara2023risk}, transportation~\cite{zhang2023empirical}, and manufacturing~\cite{gangadhara2023optimizing}, should also further managed.

\subsubsection{Vulnerability Data Quality}
The performance of security analysis tools that are developed for the software supply chain rests on the quality of the vulnerability data that is available.

\para{\challenge. Vulnerability provenance metadata}. How can we uniquely identify and track a vulnerability across the entire history\footnote{Different hosts of a library often depend on different versions of that library. Some hosts might be late or reluctant to update their dependencies.} of an ecosystem? How can we do so in bytecode or even machine code?
 How can we store, update, and access relevant metadata for each vulnerability in a standardized, machine-readable format? What are the legal and ethical considerations for the storage and availability of such potentially sensitive data? Currently, a vulnerability is assigned a unique identifier, called Common Vulnerabilities and Exposures (CVE) by a CVE Numbering Authority (CNA) and stored in vulnerability databases.
The first vulnerability database was the Repaired Security Bugs in Multics project published on February 7, 1973. Major vulnerability databases such as the ISS X-Force database, Symantec/SecurityFocus BID database, the Open Source Vulnerability Database (OSVD), and the National Vulnerability Database (NVD) aggregate a broad range of publicly disclosed vulnerabilities, including CVEs. Many SCA tools use these databases to identify known vulnerabilities in third-party components. Therefore, vulnerability databases must be updated regularly to ensure the maximum effectiveness of these tools.

However, there does not exist a standardized, machine-readable format for the associated metadata, such as the vulnerable versions, the patch, or the proof-of-vulnerability. We should develop the mechanisms needed to track a vulnerability through the ecosystem, and to swiftly update the metadata as remediation proceeds and new facts emerge. The metadata should be comprehensive, current, and trustworthy. While we believe in responsible disclosure in favor of the dependants, we should investigate the legal or ethical consequences of tracking such potentially sensitive vulnerability information at the ecosystem scale.

\para{\challenge. SCA tool benchmarking}. How do we soundly evaluate the capabilities of SCA tools~\cite{dann2021identifying}? What about domains where closed-source components play an important role, such as automotive, IoT, or blockchain? The scope of third-party artifacts included in the SCA feature databases could also heavily influence SCA capabilities~\cite{10.1145/3597926.3598143}. Though many experimental SCA tools are proposed to improve accuracy, they are mostly only validated on a limited feature dataset, i.e., by filtering open-source projects by metrics such as stars~\cite{dabic2021sampling}. This is also why many experimental SCA tools reach high accuracy but seem less satisfactory in real-world scenarios. Moreover, it is also difficult to collect a high-quality feature dataset for specific domains, such as automotive~\cite{halder2020secure}, IoT~\cite{mugarza2020security}, and blockchain~\cite{sun2023demystifying}, where closed-source artifacts play a much more important role and could compromise SCA detection if no corresponding datasets are well-established.

\subsubsection{OSS Supply Chain Governance}
In response to the escalating software supply chain security threats, the community and open-source ecosystem have rallied to implement various countermeasures aimed at mitigating risks.
However, there are many regulatory and sociotechnical challenges.
Addressing these challenges demands a collaborative effort to simplify security practices, share resources, and foster a culture of security awareness, ensuring that security enhancements do not impede the innovation and agility inherent to the open-source community.

\para{\challenge. Regulatory challenges}. To what extent software bills of material (SBOMs) are used and support cybersecurity? 
Various efforts from governments have been made to enhance the cybersecurity of software products by delineating regulations on adopting the SBOM for software products, such as the guidelines for security of the Internet of Things from ENISA~\cite{ENISA:2020}, the proposal of cybersecurity requirements for products with digital elements and amending Regulation (EU) 2019/1020 by the European Commission~\cite{CyberResilienceAct:2022}, and the United States Federal Government, per President Biden’s Executive Order 14028~\cite{UsExecutiveOrder:2021}. The executive order mandated NTIA\footnote{The United States Department of Commerce and National Telecommunications and Information Administration is an Executive Branch agency of the United States Department of Commerce that serves as the President’s principal adviser on telecommunications and information policy issues.} to publish the minimum elements of SBOMs, including the standards to be used to produce and consume SBOMs. To this end, SPDX (Software Package Data eXchange)~\cite{SPDX}, CycloneDX~\cite{cyclonedx}, and SWID (SoftWare IDentification) tags~\cite{NVDSWID} have been gradually proposed to describe a list of ``ingredients" that make up software components and related threats, such as vulnerabilities and license information. Corresponding SBOM generation tools are also integrated into security auditing tools.

However, the preliminary results in the literature suggest that the adoption of SBOMs by open-source projects is still low, even if there is an increasing trend probably due to the growing interest and pressure from major players~\cite{Nocera:2023:Icsme}. Moreover, the output of such tools also suffers from several limitations, such as representing dependencies at the level of components or libraries identified from dependency configuration files rather than at the level of code snippets (which could be manually copied from Stack Overflow or LLM-generated from license-protected source code repositories).

\para{\challenge. Sociotechnical challenges}. What will be the standards, best practices, tools, and guidelines for secure software development?
In addition to SBOMs, industry consortia, like the Open Source Security Linux Foundation (OpenSSF), have created standards, best practices, and tooling to enhance ecosystem-wide software supply chain security. For instance, the OpenSSF developed the Criticality Score~\cite{criticalityscore} for trustworthy library reuse, a set of best practices for package managers~\cite{ossfbestpractice}, guidelines for user dependency management~\cite{googlebestpractice}, and the ScoreCard project~\cite{ossfscorecard} for security threat evaluation. There are tools like OWASP~\cite{OWASPDep} and Dependabot~\cite{Dependabot} to solicit dependency updates, Sigstore~\cite{sigstore} for the verification and provenance of third-party components, SLSA~\cite{SLSA}, SPIFFE~\cite{SPIFFE} and SSDF~\cite{ssdf} as guidelines and infrastructure for secure software development and identity verification.

However, open-source projects are still reluctant to adopt these tools and guidelines, due to resource constraints, the complexity and usability of the tools, interoperability issues across diverse systems, and the need to keep pace with an evolving threat landscape. Additionally, striking a balance between rigorous security measures and maintaining development productivity poses a significant challenge.

\subsection{Beyond Memory Safety}
\label{sec:beyond-mem-safety}

Today's most critical security flaws are due to violations of memory safety. For instance, 78\% of confirmed exploited ``in-the-wild" vulnerabilities on Android devices \cite{android} and 70\% of vulnerabilities in Google Chrome \cite{chrome} are violations of memory safety. However, we see memory safety increasingly addressed  at the programming-language level. For instance, when the Android team adopted the Rust programming language for new code, the proportion of memory safety-based vulnerabilities dropped from 76\% down to 35\% \cite{android}. As memory safety vulnerabilities decrease in abundance and our security tools become more effective, attackers will focus on other types of vulnerabilities to exploit.

\para{\challenge. Emerging vulnerability types}. In the absence of memory-safety issues, which other types of software vulnerabilities exist and how can they be mitigated? How can we operationalize and detect violations of privacy or the General Data Protection Regulation (GDPR)? Memory-safety issues are conceptually easy to detect. Other types of vulnerabilities, such as injection attacks (e.g., command/code injection or deserialization attacks), data races (e.g., TOCTOU), or information leaks (e.g., side channels / information flow) require much more threat modelling from the security practitioner's point of view. In the future, we should support this modelling process.
More generally, we should design practical methods for specifying software properties~\cite{patch-specs,design-by-contract}.

When memory-safety is solved, we expect that offensive security will move on to the next low-hanging fruit. From an economical perspective, an attacker wants to maximize the likelihood of success with the least possible effort. We should empirically monitor this shift and develop the corresponding mitigations in time.

\para{\challenge. Input validation and sanitization}.
How can we make sure that adversarial inputs do not compromise the security of a software system?
If we assume that any public input to a software system is ``tainted'' and can be controlled by a malicious user, it is crucial to ensure that such input is properly validated and sanitized.
For example, a well-known class of attacks is command injection, where a malicious user injects commands in a public input, and when this input propagates through the software, some component of the software system can unintentionally execute the injected command, resulting in loss of data or leakage of secret information.
To track how much tainted data propagates through the software and to detect potentially vulnerable program points, static and dynamic program analysis techniques have been developed~\cite{clause2007dytan,tripp2009taj}.
In order to prevent the propagation of malicious inputs, it is crucial to validate (i.e., check if the input matches the expected format) and sanitize (i.e., transform the input to the expected format) the user input~\cite{balzarotti2008saner,alkhalaf2014semantic}.

However, the additional validation- and sanitization-related code also increases the attack surface and might introduce security vulnerabilities.
As the processed user input is often given as a string, this requires effective string-based program analysis \cite{stringanalysis,loops:pldi19}.
Another challenge is the fact that input validation and sanitization code is typically distributed in different parts of a software system without a clear specification of the \emph{intended} input validation and sanitization policies. Finally, new types of attacks may require changes to the existing policies, and continued modifications to the input validation and sanitization code. Due to these challenges discovering and eliminating errors in input validation and sanitization code will likely continue to be an important area of research in the future.

\para{\challenge. Sensitive data exposure}. How to detect, quantify, minimize, and eliminate the leakage of sensitive or secret data, such as customer data or cryptographic keys, due to software or hardware side-channels for the next generation of software systems?
A sensitive data exposure occurs when (properties of) secret data currently processed can be learned by observing the behavior of the processing software system. The system behavior includes side channels such as the execution time or memory/energy usage. Specifically, noninterference requires that publicly observable properties of program execution are independent of any secret values. However, the binary notion of noninterference is not entirely practical as software systems may be reasonably expected to reveal some amount of information that depends on secret values. For example, the purpose of a password checker is to disclose if the provided input matches the password (which is secret)---violating the noninterference property. Hence, there is increasing interest in quantitative information flow (QIF) which asks ``how much'' rather than ``whether'' secret information is leaked~\cite{Smi09}. The amount of information leaked is quantified using concepts such as channel capacity~\cite{val2016precisely} and Shannon entropy~\cite{BKR09,PBP17}.

Going forward, we expect many more types of side channels to be discovered. For every side channel, the various causes of leakage should be systematically identified. For instance, for a timing-based side channel, it is necessary to identify all data-dependent optimizations\footnote{Data-dependent optimizations, like speculative execution, might change the (observable) execution time depending on properties of the secret data.} in the hardware, the software, and the entire software supply chain (including the compiler~\cite{BRB20}).
In general, we need strategies for defensive programming and automated tooling to measure and minimize any possible information leakage.
One interesting direction of research is to develop automated attack synthesis techniques for both assessing the criticality of the vulnerability and also informing the mitigation strategies, where the length of the synthesized attack is inversely correlated with the severity of the information leakage~\cite{PBP17}.
We expect that both noninterference analysis and quantitative information flow analysis will continue to be important areas of research to develop techniques that identify, quantify, and eliminate information leaks due to side channels.

\subsection{Beyond Monolithic, Fully Virtual Software Systems}

Research on software security analysis has traditionally focused on monolithic programs that run on laptops, desktops, or servers. However, as our software systems become more decentralized and integrated more deeply with the physical world, we must develop new software security analysis approaches that work across large distributed systems and on proprietary hardware, when resources such as energy, time, or computing are scarce, and to avoid physical harm.

\para{\challenge. Security of distributed systems}.
How can we analyze the security of software systems (e.g., IoT, Software-as-a-Service, distributed algorithm implementations, blockchain systems, smart contracts), which are distributed across many machines and devices and often dynamically composed at runtime (e.g., as the workload changes, as components are updated, or as machines or devices become (un)available)? How can we minimize data exposure and information leaks for software components that interact via the internet?
While we believe that the following challenges apply, at least in part, to all these systems, we pick IoT as a concrete example.
The IoT revolution has been enabled by affordable and reusable embedded and cloud software platforms that integrate solutions for various design challenges.

\emph{Supply chain security}. Practically all systems use additional third-party components (for IoT, these are peripheral drivers, board support packages, and application-specific libraries).
These third-party components can represent 90-99\% of the code base. Yet, keeping up to date with issues and security releases across a diverse collection of externally-maintained components is notoriously difficult.
Hence, IoT supply chains are intricate systems that involve the production and integration of both hardware and software components, as well as the establishment of trust between different parties. The supply chain of the entire software system may not be considered as localized only within individual devices but rather as spanning across a larger system of systems, i.e., edge, fog, and cloud. The dynamic nature of these systems brings additional challenges to the supply chain security analysis. Different devices may run different (versions of)  third-party components. Going forward, improving supply chain security is critical and standardized, shareable Software or Hardware Bills of Materials (SBOM/HBOM) are a major advance towards that goal as they bring transparency to an otherwise opaque supply chain. 

\emph{Security analysis}. From an analysis perspective, distributed systems are particularly challenging for several reasons. Firstly, distributed systems may change dynamically. Computing nodes of different kinds with different versions may be added or removed elastically at any time. It is difficult to simulate this flexibility in a static analysis or in a lab/testing environment. Secondly, there is often no central agent that can be analyzed or tested. Thirdly, there are domain-specific types of vulnerabilities, like the Byzantine generals problem where decentralized parties need to arrive at consensus without relying on a trusted central party. Some security analysis questions center around the trustworthiness of the individual nodes in a distributed system. These analysis challenges require specialized approaches.

\para{\challenge. Data segregation in cloud systems}.
How do we ensure the privacy of the immense amount of sensitive data stored in cloud systems?
Software systems that have high and dynamic demands on computational resources often run on compute clouds using platforms such as Amazon Web Services (AWS), Microsoft Azure, and Google Cloud Platform (GCP). Protecting the integrity and confidentiality of sensitive user data stored in these clouds is a critical problem now, and will continue to be a critical problem in the future.
Today, \emph{access control rules} \cite{aws-iam,xacml,cancan-website,pundit-website} are explicitly and often manually specified to identify exactly who has access to what part of data, while denying unauthorized accesses.
However, errors in access control policy specifications can result in the exposure of millions of customers' private data to the public~\cite{djleak,aleak,vleak,azureflaw}.
To check the correctness of access control policies, formal verification techniques have been applied in the past~\cite{FKM05,HB08}; and in recent years they have shown promising results in practical applications~\cite{BBC18,ESL22}.

We find that access control verification is an area where the scalability of formal verification techniques seem to be sufficient for practical use.
However, many challenges remain. The capabilities of existing constraint solvers should be extended to handle the types of constraints encountered in this domain. The usability of the analysis tools should be improved by providing easy ways to specify correctness properties (such as relying on differential analysis rather than asking users to write assertions). Fully automated analysis should be achieved without generating any inconclusive results (which may require hybrid approaches that combine formal reasoning with automated testing in cases where formal reasoning is unable to prove or disprove correctness).

\para{\challenge. Cyber-physical systems} (CPS). How do we ensure the security and safety of software systems that physically interact with an evolving and incompletely perceived environment, where insecurity can cause physical damage and potentially loss of life (e.g., autonomous vehicles, IoT devices, robots, or virtual reality headsets)? CPSs are difficult analyze both physically and virtually. When testing a CPS \emph{physically} (i.e., in the actual environment), the testing procedure ma may be relatively too slow and the CPS may break or get damaged. When analyzing a CPS \emph{virtually} using static analysis (SAST) or simulation, the CPS cannot be analyzed without certain assumptions about the environment. These assumptions may or may not be true in a physical setup.

\subsection{Resilient Computing}

Widespread successful attacks on IT systems and even critical infrastructures have repeatedly proven that current common security measures are insufficient. System developers and maintainers commonly try to secure their systems by identifying, triaging, and fixing vulnerabilities. A system is commonly deemed secure when it is free of \emph{known} vulnerabilities - but what about the ones we do not know yet? Those vulnerabilities pose a strong technical debt that future attackers will happily exploit.

The only useful paradigm is, therefore, to ``assume breach''. We must design systems with the mindset that they are and will be vulnerable and, nonetheless, should be able to successfully withstand at least certain classes of common attacks, they must become attack-resilient. The guiding principle towards attack-resilience, defense in depth, has long been known, yet is at the core of too few systems we build today. Most software architectures still assume a single protection layer, which, if penetrated, puts the entire software system at risk of failure. But how to best implement multi-layered defenses?

\para{\challenge. Risk analysis}. How can we perform effective risk analysis already at the design level to judge a system's attack resilience? An analysis of attack resilience must incorporate different and complementary possible types of mitigations, ranging from process-based mitigations, such as a four-eyes principle or the use of validation tools during coding, to application-level measures, such as proper password hashing to platform-level measures, such as proper, risk-centric compartmentalization of individual subsystems.

\para{\challenge. Risk-centric compartmentalization} requires research in systems security: Currently, we can isolate subsystems essentially only by turning them into separate operating-system processes, which comes with huge computational and maintenance overhead. But new technologies are on the horizon.
WebAssembly, for instance, allows one to sandbox untrusted parts of applications with relatively low overhead. Existing compilers allow one to create WebAssembly for C/C++ components.
Other opportunities to isolate computations within applications may be yielded by using hardware support such as Intel’s Memory Protection Keys, ARM Memory Tagging and CHERI~\cite{cheri}. These are relatively low-level features, yet researchers can test their efficacy as security barriers when properly integrated into programming languages and environments. Further, GraalVM, in a feature still under development, now supports the proper isolation of code executing within the same operating system process, currently in polyglot but soon also in single-language settings.

\subsection{Emergent Behaviors and Vulnerability Composition}

Computer security research is sometimes seen as an arms race where the advances in defensive techniques that eliminate vulnerabilities lead to advancements in offensive techniques that result in new types of attacks, which then inspire new defensive strategies, resulting in a continuous cycle of security measures and counter-measures. In this race, the defensive techniques must be in the lead to succeed in securing computer systems.
An increasingly critical set of offensive techniques relies on emergent behaviors and compositions of vulnerabilities to create exploit chains that achieve the adversarial goals of the attacker.

The concept of ``weird machines'' characterizes exploit generation as a task of programming a weird machine, where the instructions of the weird machine are vulnerabilities or unintended behaviors that take the system to an unintended ``weird'' state~\cite{dullien11:weird,bratus12:weird}. In this characterization, exploit chaining corresponds to writing a weird machine program that achieves the attacker's goal. Defending against this type of sophisticated attack strategy requires a defensive strategy that not only detects emergent behaviors but also analyzes their compositions.

\para{\challenge. Identifying emergent behaviors across layers of abstraction}.
How can we identify emergent behaviors that cut across different layers of software, for example, emergent behaviors that involve a combination of vulnerabilities in the firmware, operating system, and application code?
Software is built at different layers, where a layer exposes an interface providing an abstraction to the next layer. However, in many cases, these interfaces lead to unintended behaviors due to misuse, under-specification, or ambiguous specification of the interface constraints. Although analysis of software interfaces has been investigated in the past, novel techniques that can identify unintended/emergent behaviors across multiple layers of abstraction that involve multiple interfaces are needed to assess the security of the whole software system. This will require innovative analysis techniques that can track the behavior of the system across multiple layers of abstraction.

\para{\challenge. Discovering and defending against exploit chains}.
How can we analyze collections of vulnerable or unintended behaviors to prevent attacks that chain multiple vulnerabilities?
An unintended behavior in a computer system may not be a security vulnerability by itself. However, an attacker who is aware of a set of unintended behaviors can combine them to generate a programmable collection of adversarial operations (which correspond to instructions of a weird machine) and then write programs that chain these operations. In order to defend against this type of attacks, we have to first discover them. Exploit chains  can be seen as a composition of emergent/unintended behaviors. Techniques that can automatically compose emergent/unintended behaviors can lead to techniques that can automatically search for exploit chains. Once the exploit chains are discovered, mitigation strategies can be developed to identify the most effective ways to eliminate them.

\subsection{Security Education}
\para{\challenge. Security as a mandatory course}.
There is a gap between the need for personnel skilled in software security and the availability of that personnel~\cite{SecurityMagazine:2021}. Computer science students still graduate with scarce or no secure programming knowledge~\cite{almansoori2020secure}. It becomes necessary to provide adequate security training to the next generation of developers. This will be possible by (i)~improving students’ engagement in producing secure software (traditional, ESs, CPS, and IoT systems), (ii)~helping students acquire security skills and knowledge on the use of development methods for secure development, (iii)~training students on Dynamic and Static Analysis Security Tools (DASTs and SASTs, respectively), and (iv)~educating students on the perils related to the use of AI-generated code and code snippets from Questions \& Answers forums. In continuous education, the education of ``security champions'' within companies has shown great successes. How can this be leveraged on a global scale?

\para{\challenge. Integrate security tool usage into software engineering courses}.
While introducing some security concerns in the source code is unavoidable (e.g., because a certain vulnerability is still unknown), others can be fixed before the delivery of software systems (or even developers could avoid their introduction). Educators should train the next generation of developers so that they identify known security concerns and then fix them before the delivery of software systems (e.g.,~\cite{NoceraRFS23,NoceraRFS24}). This could be achieved by training students on DAST and SAST tools that are well known in both the academy and industry. Specifically, DAST tools analyze software systems at run-time, while SAST tools examine the code of software systems without executing them. There are three scenarios in which developers can leverage SAST tools: \textit{(i)}~while developing code, by highlighting the presence of security concerns directly in IDEs (Integrated Development Environments); \textit{(ii)}~within a Continuous Integration pipeline, which could make a build fail if the code is not compliant with given security rules (e.g., the committed code must not contain critical vulnerabilities); and \textit{(iii)}~during a code review. A DAST tool could be used in the last two scenarios together with a SAST tool.

\subsection{Law and Policy Making}
\textbf{\challenge. Law and policy making}.
Apart from the sociotechnical challenges, how can we ensure that our critical software infrastructure is robust against cyberattacks, we need effective policies and and regulation? How can we balance legal and technical aspects for critical as well as non-critical systems, to both empower users and secure the interactions? Should the vulnerability disclosure process or supply chain security practice be regulated, and if so how?
While various industries where software plays a key role are already strictly regulated (such as safety-critical domains like aerospace), software products themselves lack such legal safeguards.
We have previously discussed regulations for the software supply chain in some domains like the IoT, including the need to accompany a software product with a Software Bill of Materials (SBOM) \cite{ENISA:2020,CyberResilienceAct:2022,UsExecutiveOrder:2021}. However, much more work is needed with legal scholars, economics researchers, policy makers, and software vendors to create an effective legal framework and incentive system.
It is important for software researchers and practitioners to engage in shaping the legal framework to ensure that appropriate cybersecurity practices are applied and lessons are learned after each cybersecurity incident.

\section{Conclusion}

In this paper, we have outlined our vision of software security for the software systems of the future. Specifically, we have drawn a research roadmap towards 2030 and beyond. Starting from state-of-the-art recent advances in software security, we have identified concrete challenges and opportunities for the security analysis of the software systems of the future and provide specific directions of research for the software engineering community. We have also discussed open challenges and opportunities and presented a long-term perspective for software security in the context of software engineering.

%

\bibliographystyle{ACM-Reference-Format}
\bibliography{references}

\appendix

\end{document}